\documentclass{jpsj2}

\title{Business Cycle and Conserved Quantity in Economics}
\author{
Masa-aki Taniguchi$^{1}$\thanks{E-mail address: mass@vega.aichi-u.ac.jp},
Masako Bando$^{1}$,
and Akihiro Nakayama$^{2}$}
\inst{%
$^{1}$Aichi University, Miyoshicho, Aichi 461-8641, Japan. \\
$^{2}$Meijo University, Nagoya 468-8502, Japan.
}

\abst{%
We propose a dynamical model for business cycle based on an optimal DI model.
In the model there exists a conserved quantity, which corresponds to
the total energy in a dynamical system. We found that the business cycle 
with the period $6 \sim 7$ years is nicely reproduced, since the model 
predicts a periodic motion in the conservative system. 
}
\kword{%
business cycle, econo-physics, economic fluctuations
}

\date{January 08, 2007}

\begin{document}
\maketitle

\section{Introduction}

\begin{figure}[htbp]
\centerline{
\includegraphics[height=6cm]{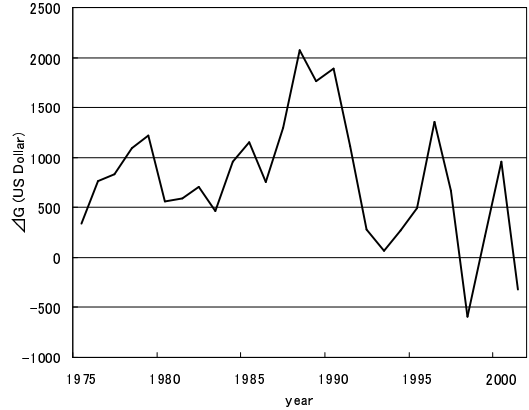}
}
\caption{Observed data of $\Delta G(i)$ which is calculated from the GDP
per capita (constant 1995 US dollar) in Japan\cite{wdi}.}
\label{dg}
\end{figure}

Time dependence of real gross domestic product (GDP) usually consists
of the fluctuations under the long term growth background.
In order to see the profile and the fundamental structure of such fluctuation
more clearly, we proposed to extract $\Delta G(i)=GDP(i)-GDP(i-1)$,
since we are free from the background of increasing function.
$\Delta G(i)$ commonly shows a kind of cycle repeating depression
and prosperity, and it is called ``business cycle" (Fig.~\ref{dg}).
Although such business cycle may be classified
into several types according to characteristics, especially its period:
the Kitchin inventory cycle, the Juglar fixed investment cycle, the Kuznets
infrastructural investment cycle and the Kondratieff wave, those have
the period $3 \sim 5$ years, $7 \sim 11$ years, $15 \sim 25$ years 
and $45 \sim 60$ years respectively.

The origin of such business cycle can be divided into external and
internal sources. Many authors have proposed various models to explain
why the business cycle occurs\cite{sam1, hicks, inada_uzawa,
arn, puu, sam2, kaldor, goodwin}.
From the physical view point, we are interested in the internal origin
where the business cycle is caused by some dynamics.

In the previous paper we have introduced another dynamical quantity,
DI (Diffusion Index) in addition to  $\Delta G$, which determine the dynamics
of the economic system. We have proposed a model, an optimal DI model,
which reproduces a fluctuating behavior\cite{our07}.
Our model has nice features, explaining oscillations of  $\Delta G$ and $DI$
and the typical tendency of fluctuation:
$\Delta G(t)$ increases when the corresponding DI is smaller than optimal DI,
and decreases when DI is larger than optimal DI.

However we cannot reproduce typical business cycle in the phase space
from our previous model.
In this paper we modify our model to explain the business cycle.
In the new model, the difference equation used in the previous model is
replaced by the differential equation.
We find that there is a conserved quantity in the system described by
this differential equation.

After making a brief review of the original optimal DI model in section \ref{sec2}, 
we modify the model and introduce the total energy in section \ref{sec3}.
We investigate the properties of the modified model in section \ref{sec4}.
The section \ref{sec5} is devoted to the investigation of the variation of 
the parameters of this model. We compare the energy and the period to 
the real data in section \ref{sec6}.
We make the concluding remarks in the last section.

\section{The Original Optimal DI model}\label{sec2}

First we make a brief review of the original optimal DI model\cite{our07}.
The dynamical quantity DI is found in ``Tankan", which is announced by
Bank of Japan as the Short-term Economic Survey of Enterprises in Japan.
It is obtained from the the ratio of the number of answers "Yes/No"
(denoted by $Y$/$N$) to the question "Is your business good?"
\[
DI=100\frac{Y-N}{Y+N}.
\]

In general, the dynamical quantities $DI(t)$ and $\Delta G(t)$
are continuous functions of time $t$.
However we have used the variables $DI(i)$ and $\Delta G(i)$ instead of
$DI(t)$ and $\Delta G(t)$, because
the data points of $DI(t)$ and $\Delta G(t)$ are given annually.


We have constructed an optimal DI model,
which is analogous to the optimal velocity model of traffic flow\cite{traffic},
by introducing the ``optimal DI" (ODI) function
\begin{equation}
ODI(\Delta G)=A + B \tanh(C(\Delta G-D)),
\end{equation}
where A, B, C, and D are constant parameters.
The ODI function can be fixed from the observed data.
In the previous paper, we adopted the parameters
\begin{equation}
A = -5, B = 55.3, C = 6.28 \times 10^{-4}, D = 880.
\label{params1}
\end{equation}
The dynamical equations are represented as a set of the difference
equations as follows:
\begin{eqnarray}
 DI(i+1)-DI(i) &=& a (ODI(\Delta G(i))-DI(i))
\label{disc1}\\
 \Delta G(i) &=& b DI(i) + c,
\label{disc2}
\end{eqnarray}
where $a$, $b$, and $c$ are constants.

These equations come from the following consideration.
The presidents of companies make their decision $DI(i)$ of business activity.
Equation (\ref{disc1}) represents how the presidents modify their
decision in order to adapt their $DI$  to $ODI$:
If $DI(i)$ is smaller than $ODI$, $DI(i)$ is modified to larger one,
and vice versa.

$DI$ represents the tendency of the presidents of companies
to accelerate or decelerate their production rates.
Therefore the economic growth $\Delta G(i)$ is a reflection of $DI(i)$,
namely $\Delta G(i)$ is a function of $DI(i)$.
In Eq.(\ref{disc2}),
we suppose $\Delta G(i)$ is a linear function of $DI(i)$.
The parameters are determined from the observed data,
$b = 23.8,\ c = 980$.

\section{Modification of ODI model}\label{sec3}

In the previous model, we have adopted a linear function (\ref{disc2}).
However we find that the correlation between $[\Delta G(i)+\Delta G(i-1)]/2$
and $DI(i)$ is stronger than that between $\Delta G(i)$ and $DI(i)$
In fact the correlation coefficient of the former is $0.94$,
and that of the latter is $0.85$.
We modify the second equation (\ref{disc2}), and our model is expressed by
a set of the following dynamical equations,
\begin{eqnarray}
 DI(i+1)-DI(i) &=& a [ODI(\Delta G(i))-DI(i)],
\label{disc1b}\\
 \Delta G(i+1)+\Delta G(i) &=& 2bDI(i+1)+2c,
\label{disc2b}
\end{eqnarray}
where
\begin{equation}
b = 23.6,\quad  c = 969.
\label{params2}
\end{equation}

From Eq.~(\ref{disc1b}) and  Eq.~(\ref{disc2b}), we obtain the following equation,
\begin{eqnarray}
\Delta G(i+1)+a \Delta G(i)+(a - 1)\Delta G(i-1)
= 2abODI(\Delta G(i))+2ac.
\label{eq6}
\end{eqnarray}
Here we note that the economic system in the above model is expressed
in terms of a single dynamical variable $\Delta G(i)$.
Hereafter we denote $x(i)$ instead of $\Delta G(i)$.

By solving Eq.(\ref{eq6}) numerically, we find that the behavior of
$x(i)$ is dependent of the parameter $a$.

\begin{enumerate}
\item[(1)]  $a<2$ \\
The behavior of $x(i)$ is similar to that of the original ODI model\cite{our07},
namely $x(i)$ tends to a fixed point irrelevant to the initial condition.
\item[(2)]  $a=2$ \\
This is a special case and we find that $x(i)$ shows a periodic behavior.
\item[(3)]  $a>2$ \\
Irrelevant to the initial condition, $x(i)$ tends to infinity.
\end{enumerate}
\vspace{\baselineskip}

Now we concentrate on the case $a=2$ and rewrite Eq.(\ref{eq6}) as follows:
\begin{equation}
x(i+1)-2x(i)+x(i-1) = 4[bODI(x(i))-x(i)+c].
\label{eqm}
\end{equation}
The left hand side of Eq.(\ref{eqm}) can be regarded as the second derivative of
the continuous function $x(t)$. So we replace the difference equation (\ref{eqm})
by the differential equation
\begin{equation}
{d^2x(t) \over dt^2}= 4[\ bODI(x(t))-x(t)+c\ ]
\label{eqm_con}
\end{equation}
as a fundamental equation.

The right hand side Eq.~(\ref{eqm_con}) is a force, which
is written in terms of $x(t)$ only. This indicates that
our model represents a conservative system. Then we can define
the conserved quantity "total energy"
\begin{eqnarray}
E &=& {{1 \over 2}\left({dx \over dt}\right)^2 + V(x)}, \label{totenergy}
\\
V(x) &=& -\int 4[\ bODI(x)-x+c\ ]\ dx, \label{potenergy}
\end{eqnarray}
where $V(x)$ is the potential energy of this system.
If we use the ODI function
\begin{equation}
ODI(x)=A + B \tanh(C(x-D)),
\end{equation}
the potential energy is calculated as
\begin{eqnarray}
V(x) = -4{bB \over C}\log(\cosh(C(x-D)) + 2(x-bA-c)^2 + const.
\label{pot}
\end{eqnarray}
In the following discussion, the constant in Eq.(\ref{pot}) is set to zero
for convenience.

\section{Properties of the model}\label{sec4}

In this section we investigate the properties of our model.
The dynamical variable of the model is $x(t)$ only,
in terms of which the system is expressed by the following Hamiltonian.
\begin{eqnarray}
H &=& {{1 \over 2}\left({dx \over dt}\right)^2 + V(x)}, \label{hamiltonian}
\\
V(x) &=& -4\beta\log(\cosh(C(x-D)) + 2(x-\gamma)^2, \label{potenergy1}
\end{eqnarray}
where $\beta \equiv bB/C$ and $\gamma \equiv bA+c$.
The parameters $A$, $B$, $C$, $D$, $b$ and $c$ are the same as
Eqs.(\ref{params1}) and (\ref{params2}), which have been determined
in Section \ref{sec2}.
Hereafter we take the unit of $x(t)$ as $10^3$ dollars;
$x(t) =\Delta G(t) \times 10^{-3}$ for convenience.
Then the values of parameters of the above equations are converted to
\begin{equation}
\beta=2.08,\ \gamma=0.851,\ C = 0.628,\ D = 0.880.
\end{equation}

\begin{figure}[htbp]
\centerline{
\includegraphics[height=6cm]{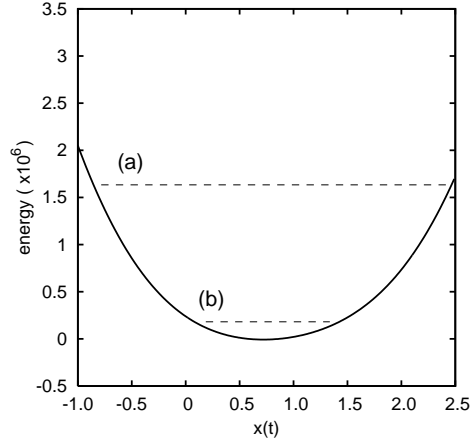}
}
\caption{Solid line represents the potential energy (Eq.~(\ref{hamiltonian})).
Two dashed lines (a) and (b) represent the energy calculated from the real GDP
data in $1974$ and $1979$ respectively.
}
\label{potential0}
\end{figure}

Figure \ref{potential0} shows the shape of the potential energy.
The total energy, which is
conserved quantity, is a sum of this potential energy and kinetic energy.
However in realistic situation irregular changes of economy
such as the Depression in 1929 happen to occur.
In our model, such irregular changes correspond to the external forces.
We shall discuss on this point in section \ref{sec6}.
The total energy jumps to a different value due to such effects, and accordingly
the total energy evaluated from the observed data is not really constant.
For example, the dashed line (a) in Fig.~\ref{potential0} represents the total energy
evaluated from the
data in 1974 (high energy case), and (b) represents that in 1979 (relatively
low energy case).
From this figure, we find that $x(t)$ oscillates between $-800$ and $2500$
in the case (a), and oscillates between $200$ and $1300$ in the case (b).

\begin{figure}[htbp]
\centerline{
\includegraphics[height=6cm, width=6cm]{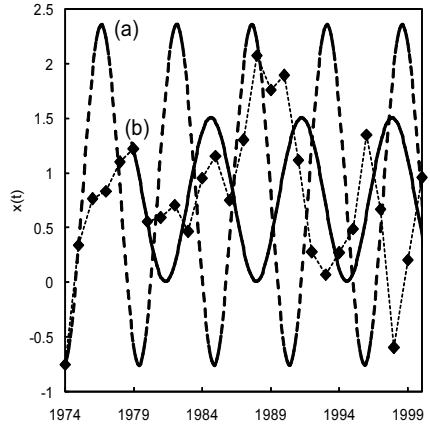}
}
\caption{Behaviors of $x(t)$ using the data in (a) $1974$ (dashed line)
and (b) $1979$ (solid line) as the initial condition. As a reference, we add
the observed data (dotted line with diamond marks).}
\label{pred_dg}
\end{figure}

In order to investigate the motion of $x(t)$, we solve Eq.~(\ref{eqm_con})
numerically in the above typical cases.
Figure \ref{pred_dg} shows the results of the simulation using the data in
(a) $1974$ and (b) $1979$ as the initial condition. The total energy
defined in Eq.(\ref{hamiltonian}) is (a) $E=1.57$ in $10^6$ unit and
(b) $E=0.12$. The period of the motion
is $5.5$ and $7$ years in the case (a) and (b), respectively.
To guide the eye, we plot the real data points.

\begin{figure}[htbp]
\centerline{
\includegraphics[height=6cm, width=6cm]{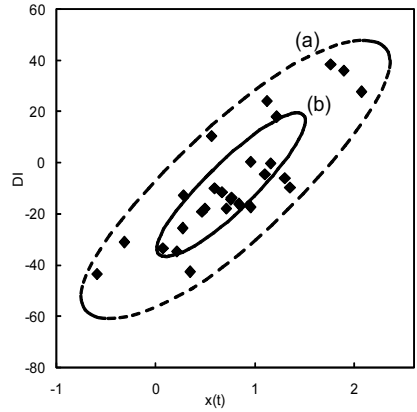}
}
\caption{Trajectories in the phase space $(x(t),DI(t))$. Dashed and solid curves
represents trajectories starting from the data in (a) $1974$
and (b) $1979$. Diamond marks show the observed data.}
\label{phase_space_diagram}
\end{figure}

In order to understand the behavior of business cycle more clearly,
we show the trajectories in the phase space $(x(t),DI(t))$
(Fig.\ref{phase_space_diagram}).
This phase space plays an essential role in the original ODI model\cite{our07}.
Here $DI(t)=(1/2b)[x(t)+x(t-1) -2c]$ (see Eq.(\ref{disc2b})).
As a reference, we add the observed data.
The periodic motion of $x(t)$ with corresponding $DI(t)$ is visualized
by close circles in Fig.\ref{phase_space_diagram}.
It is found that the higher the total energy is, the larger the circle becomes.

Here let us investigate the energy dependence of the period of motion.
We have seen in Fig.\ref{potential0} that the global shape of the potential
is approximately of the harmonic oscillator type.
Indeed, if $x(t)$ is large, the second term in Eq.(\ref{potenergy1}) dominates
and then the potential becomes that of a harmonic oscillator,
$V(x) \rightarrow 2x^2$. This is independent of the parameters $\beta$,
$\gamma$, $C$ and $D$.
Thus, for large $x(t)$, the equation of motion (\ref{eqm_con}) reduces to
the following simple form
\begin{equation}
{d^2x(t) \over dt^2}= -4x(t).
\label{eq_simple}
\end{equation}
If the system is described by this equation, $x(t)$ moves periodically
with the period $\pi$, independent of the total energy.
The larger the total energy is, the wider the region becomes where
the equation of motion of $x(t)$ is approximated by Eq.~(\ref{eq_simple}).
In such case, the period is determined mainly by the behavior of large $x(t)$,
even if the potential is deviated from the
harmonic one for smaller $x(t)$. On the other hand,
if the total energy is small, the effect from the
potential deviated from the harmonic one becomes important.
We have found by numerical calculation that this effect makes the period
longer (see Fig.\ref{pred_dg}).

\begin{figure}[htbp]
\centerline{
\includegraphics[width=6cm]{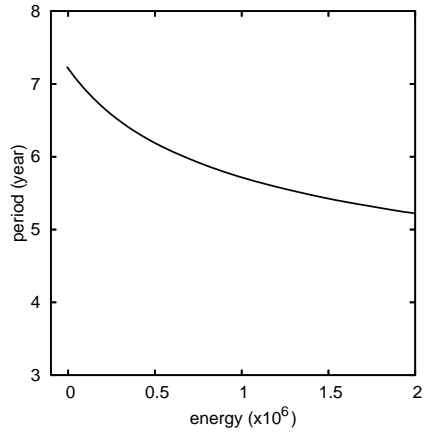}
}
\caption{The period of this system is shown. The period is a monotonically
decreasing function of $E$, which tends to the value $\pi$.
 }
\label{period1}
\end{figure}


To see the energy dependence of the period, we calculated the
period numerically. Figure \ref{period1} shows that the period is a
monotonically decreasing function of the energy $E$, and tends to
the value $\pi$.
The reason of this result is as follows.
The first term of Eq.(\ref{potenergy1}) makes the potential flatter,
and consequently the period becomes longer.
Note that this result depends on the choice of parameters
as we shall see in the next section.

\section{Variation of parameters}\label{sec5}

In this section we discuss the variation of the parameters $\beta$, $\gamma$,
$C$ and $D$ to see how the results of the previous section depend on those
parameters. The parameters of the ODI function
are determined from the observed data. However we have some freedom in
the choice of the ODI function, because the data are widely scattered.

\begin{figure}[htbp]
\centerline{
\includegraphics[height=6cm]{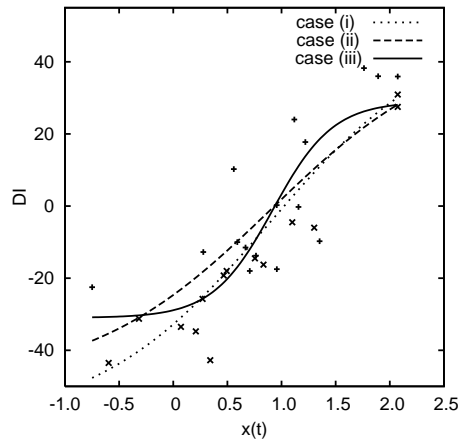}
}
\caption{ODI functions in three cases.
The dotted curve of case (i) is the original ODI function used in the
previous section. The dashed and solid curves labeled as case (ii) and (iii)
are a gentle and a steep ODI function, respectively.}
\label{odi2}
\end{figure}

As examples,  we pick up three typical choices of the ODI function from the
same observed data (Fig.\ref{odi2}). The dotted curve labeled as ``Case (i)'' is
the original ODI function used in the previous section. In the ``Case (ii)''
and ``Case (iii)'', we choose a gentle and a steep ODI function, respectively.
The parameters in each case are summarized in Table \ref{table1}.

\begin{table}[htb]
\caption{Parameters of the ODI function in three cases (i), (ii) and (iii).}
\begin{tabular}{|l|c|c|c|c|c|}
\hline
	& $\beta$ & $\gamma$ & $C$ & $D$ & \\
\hline
case (i) & 2.08 & 0.851 & 0.628 & 0.880 & $(A=-5,\ B=55.3)$ \\
case (ii) & 1.80 & 0.945 & 0.600 & 0.900 & $(A=-1,\ B=48.0)$ \\
case (iii) & 1.13 & 0.945 & 1.800 & 0.920 & $(A=-1,\ B=30.0)$ \\
\hline
\end{tabular}
\label{table1}
\end{table}

\begin{figure}[htbp]
\centerline{
\includegraphics[height=6cm]{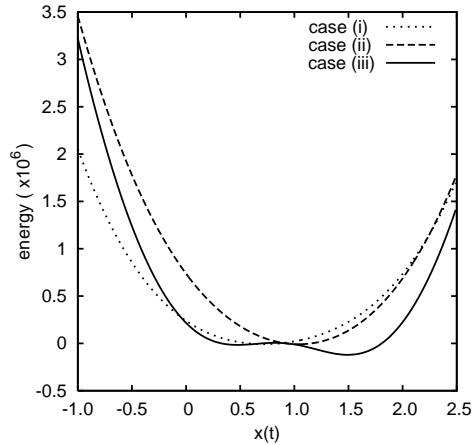}
}
\caption{Potential energies in three cases. Dotted, dashed and solid curves
represent the potential energy E in the cases (i), (ii) and (iii), respectively.}
\label{potential}
\end{figure}

Figure \ref{potential} shows the potential energy in each case.
The shape of the potential energy in the case (ii) is similar to that of the
original one, which has a local minimum. However for the steep case (iii),
the potential energy has two local minimums, which is a so-called
winebottle type.

\begin{figure}[htbp]
\centerline{
\includegraphics[width=6cm]{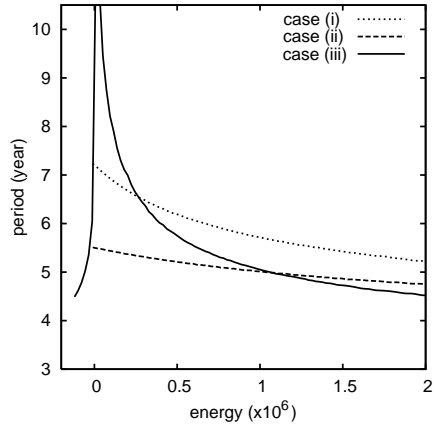}
}
\caption{Periods of the motion in three cases. Dotted, dashed and solid curves
represent the periods in the cases (i), (ii) and (iii), respectively.}
\label{periods}
\end{figure}

We calculate the period of the system numerically in each case.
The results is shown in Fig.~\ref{periods}.
In the case (ii), the period is a monotonically decreasing function of energy $E$,
as well as the original case (i). 
The period in the case (iii), on the other hand, blows up at $E=0.0061 \times 10^6$,
which is the energy at the local maximum of the winebottle potential in Fig.~\ref{potential}.

Let us explain how the motion of $x(i)$ changes as the energy increases.
If the energy takes the lowest value (at the bottom of the potential),
$x(t)$ is constant.
If the energy becomes a little larger than the minimum of the potential, $x(t)$ moves
periodically with the period $\sim 5.5$ years for the case (i), and $\sim 7$ years 
for the case(ii). At much higher energy, the period becomes shorter, and tends to 
$\pi$ in the high energy limit. (This is also common to the case (iii). )

In contrast, the period blows up in the case (iii), 
if the total energy approaches to the value at the local maximum 
($E=0.0061 \times 10^6$). 
The business cycle with the very long period such as Kondratieff wave
might correspond to this situation.

Here we comment on the case (iii). There exists a single solution if the
energy is between $-0.121$ (global minimum) and $-0.016$
(local minimum). However, for the energy between $-0.016$
and $0.0061$ (local maximum), there are two solutions for a given energy:
one has larger $x(t)$ and the other has smaller $x(t)$.
It depends on the initial condition which solution is chosen.
We note that the periods of the motion for these two solutions are
different from each other. But the difference is quite small, and we cannot
distinguish them so far as from Fig.~\ref{periods}.
Once a solution is chosen, $x(t)$ maintains the periodic motion
described by the solution and never jumps to the state described 
by the other solution is never realized. 
In the realistic situation, however, there always exists a fluctuation in $x(t)$
caused by irregular changes of economy. Then the system can transit
from one state to the other, if there arise sufficiently large fluctuations.

\section{Energy and period in the real system }\label{sec6}

\begin{figure}[htbp]
\centerline{
\includegraphics[height=6cm]{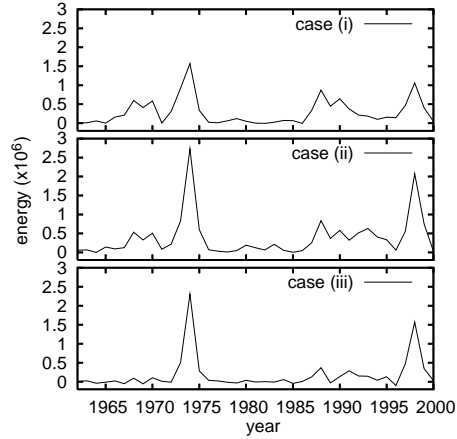}
}
\caption{Total energy E calculated from the real GDP data. }
\label{energy}
\end{figure}

Our model is a conservative system and the total energy $E$ is invariant.
The energy changes when the external force acts, that is, the economy
changes irregularly. In order to see this, we calculate the total energy
by using the real GDP data. Figure~\ref{energy} shows the plot of the energy 
calculated from the data of each year in three cases, (i), (ii), and (iii).
We can see a common behavior in all cases. There are two distinct peaks in
$1974$ and $1998$, and two broad hills around $1968$ and $1990$. 
For other years, the energy seems to be almost constant. 
It depends on the choice of the parameters of ODI function 
whether we can assume the existence of the conserved quantity or not. 
For example, in the case (iii), the energy seems to be constant in almost 
all years except two peaks.

The positions of these peaks and hills coincide with the years of 
big changes of Japanese economy.
We list the years of the big changes of Japanese economy in Table~\ref{tab0}. 
The first peak corresponds to the so-called "first oil shock" and the second peak, 
"Heisei depression". The first and the second broad hills correspond to "Izanagi boom" 
and "bubble" economy respectively. 
We find that the total energy for Japanese economic system
calculated from observed data shows almost constant except for some
points. In Fig.~\ref{energy} sharp peaks correspond to depressions and 
the broad hills correspond to booms. 

\begin{table}[htbp]
\caption{Big change of Japanese economy}
\label{tab0}
\centering{
\begin{tabular}{|l|l|}
\hline
years & economic event \\
\hline
1965-70 & Izanagi boom \\
1974 & first oil shock \\
1979 & second oil shock  \\
1986-91 & bubble economy \\
1997-98 & Heisei depression \\
\hline
\end{tabular}
}
\end{table}

The total energy increases both in booms and depressions. 
Basically we assume the total energy is an invariant quantity and its change 
arises from the external forces. Therefore the change of the total energy
can be used as an index of the change of the economic situation.

\begin{figure}[htbp]
\centerline{
\includegraphics[height=6cm]{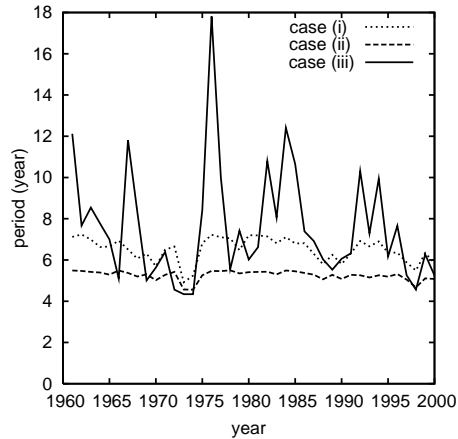}
}
\caption{The period of the motion of $x(t)$ calculated from the real GDP
data. Dotted, dashed and solid lines are the periods in the case (i),
(ii) and (iii), respectively.}
\label{period}
\end{figure}

As a reference, we also show the periods of business cycle for three cases 
in Fig.~\ref{period}. The periods are obtained from the relation between the 
period and the total energy (Fig.\ref{periods}) and the observed value of
total energy (Fig.\ref{energy}).
The period is roughly $5 \sim 8$ years in the case (i) and (ii).
On the other hand, in the case (iii), the period is $5 \sim 18$ years.
It is known that the period of the Juglar fixed investment cycle
is $7 \sim 11$ years. Hence it seems better to take the gentle 
ODI function as in case (i) and (ii).

\section{Concluding Remarks}\label{sec7}

The model which we have introduced in this paper is turned to
be a model which describes a conservative system, having a
Hamiltonian with a single dynamical variable $x(t)$.
The economic system is described in terms of a point particle moving 
in the harmonic like potential. The dynamical variable $x(t)$ moves periodically 
and the total energy $E$ of this system is conserved.
However, the observed data of GDP in Japan shows that the total energy $E$ 
is not always constant. At several points $E$ takes very large values  
as seen in Fig.~\ref{energy}. Such sudden changes correspond to 
"first oil shock", "Heisei depression", "Izanagi boom" and "bubble" economic 
events. From the viewpoint of our model, these events can be interpreted 
as consequences of some external forces, which cannot be predicted 
from our dynamical model. We can use the total energy as a sort of the 
index of such economic events. 

The observed data show that there exist the business
cycles in the economies of almost all modern countries in the world.
It is a quite general feature of economic system,
and has long been one of the most interesting questions how
to explain such business cycles.
Our model naturally explain the origin of the business cycle by the 
periodicity of $x(t)$. 
We also find the relation between the total energy and the period of
business cycle. A constant period of business cycle is a result of 
the energy conservation.
In our model the business cycle with the period $5 \sim 7$ emerges (case (i) and (ii)).
This cycle may be identified as the Juglar fixed investment cycle.
By choosing another parameters, the period of business cycle can be made 
much more longer (case (iii)). 

The behavior of the total energy calculated from the real GDP data
has an interesting feature. The energy increases quickly by irregular
events, but the effects are washed out immediately and
the energy returns to the value before the events. 
This feature suggests that our model extracts only a conservative
nature of the economic system. To construct a more realistic model 
the dissipation term may be necessary. It is our future work.

It would be interesting tasks to compare the total energy and the business
cycle with those of other countries. In real world, the economic system
of each country is affected by other countries. It is also interesting
problem to investigate the interaction among many economic systems
and their collective dynamics.

\section*{Acknowledgement}
The authors thank K. Nakanishi for helpful discussions. 
This work was partly supported by a Grant-in-Aid for Scientific Research (C) 
of the Ministry of Education, Science, Sports and Culture of Japan (No.18540409).

\end{document}